\documentclass[prb,twocolumn,showpacs,superscriptaddress,preprintnumbers,amssymb]{revtex4}
\usepackage{graphicx}
\usepackage{dcolumn}
\usepackage{bm}
\usepackage{wasysym}

\newcommand{\beq}{\begin{equation}}
\newcommand{\eeq}{\end{equation}}
\newcommand{\beqn}{\begin{eqnarray}}
\newcommand{\eeqn}{\end{eqnarray}}

\begin{document}

\title{Wave Functions of Bosonic Symmetry Protected Topological Phases}

\author{Cenke Xu}

\affiliation{Department of Physics, University of California,
Santa Barbara, CA 93106}

\author{T. Senthil}

\affiliation{Department of Physics, Massachusetts Institute of
Technology, Cambridge, Massachusetts 02139}

\begin{abstract}

We study the structure of the ground state wave functions of
bosonic Symmetry Protected Topological (SPT) insulators in 3 space
dimensions. We demonstrate that the differences with conventional
insulators are captured simply in a dual vortex description. As an
example we show that a previously studied bosonic topological
insulator with both global U(1) and time-reversal symmetry can be
described by a rather simple wave function written in terms of
dual ``vortex ribbons". The wave function is a superposition of
all the vortex ribbon configurations of the boson, and a factor
$(-1)$ is associated with each  self-linking of the vortex
ribbons. This wave function can be conveniently derived using an
effective field theory of the SPT in the strong coupling limit,
and it naturally explains all the phenomena of this SPT phase
discussed previously. The ground state structure for other 3d
bosonic SPT phases are also discussed similarly in terms of vortex
loop gas wave functions. We show that our methods reproduce known
results on the ground state structure of some 2d SPT phases.

\end{abstract}

\date{\today}

\maketitle

\section{Introduction}

The disordered ground states of strongly interacting quantum
many-body systems can have much richer structures compared with
classical disordered states. The quantum richness of a system is
encoded in the entanglement of its ground state wave function, and
without assuming any symmetry of the Hamiltonian, the ground state
wave function of a quantum many-body state can have long range
entanglement, which implies that the system has a ``topological
order". In the last few years, motivated by the discovery of free
fermion topological insulators protected by time-reversal
symmetry~\cite{kane2005a,kane2005b,bernevig2006,moorebalents2007,fukane,roy2007},
it was realized that short range entangled state can still be
fundamentally distinct from trivial product states, as long as the
system preserves certain global symmetry $G$. These nontrivial
quantum disordered phases with short range entanglement are called
``symmetry protected topological" (SPT) phases.  They are
separated from the trivial product state through sharp quantum
phase transitions in the bulk, either continuous or first order.
In space dimension $d = 1$ the Haldane spin chain provides an old
and nice example of an SPT phase~\cite{haldane1,haldane2}. It has
a bulk gap and no fractional excitations but nevertheless has
dangling symmetry protected spin-$1/2$ moments at the
edge~\cite{affleck1987,kennedy1990,hagiwara1990,ng1994}. The
Haldane chain thus provides an early example of an interacting
topological insulator.

In this paper we are mainly concerned with three dimensional
symmetry protected topological insulators of bosons/spin systems.
A formal mathematical classification~\cite{wenspt} of SPT phases
based on group cohomology allows a number of such phases to exist
(depending on the global symmetry) but sheds little light on the
physical properties. The latter have been discussed recently in
Ref.~\onlinecite{vishwanathsenthil}. A characteristic feature of
all SPT phases is the presence of non-trivial surface states at
the interface with a trivial insulator. Indeed though the bulk is
gapped and has no fractional excitations or topological order,
such an interface cannot be in a trivial insulating state.
Ref.~\onlinecite{vishwanathsenthil} described the effective
surface theory for a number of three dimensional bosonic
topological insulators and determined the structure of the allowed
non-trivial phases. These either spontaneously break the defining
global symmetry or if gapped have surface topological order.
Exotic symmetry preserving gapless states were also shown to be
possible. A key feature is that the surface effective field theory
realizes symmetry in a manner not possible in strictly two
dimensional systems. Bulk topological field theories and effective
field theory descriptions have also been provided.

In this work we will study the structure of the ground state wave
function of various such 3d bosonic SPT insulating phases with
global $U(1)$ and time reversal ($Z_2^T$) symmetries. The
differences with conventional Mott insulators are conveniently
captured in a dual description in terms of closed vortex loops. In
Mott insulating phases (conventional or topological) the vortex
loops have proliferated and the ground state wavefunction can be
described as a vortex loop gas (see Sec. \ref{wftrbmi}). We show
that when compared with the conventional insulator  this vortex
loop gas has extra phase factors depending on the topology of the
vortex loop configuration. We demonstrate that these wave
functions simply capture all the major phenomena of the SPT
phases, both in the bulk and at the boundary, that were discussed
in Ref.~\onlinecite{vishwanathsenthil}.  As a key example, in Sec.
\ref{wf3dtbi} we discuss a non-trivial SPT phase with symmetry
either  direct ($U(1) \times Z_2^T$) (as is appropriate for a spin
model realization of an interacting boson system) or semidirect
($U(1) \rtimes Z_2^T$) product. Here the vortex lines should be
viewed as ribbons with a non-zero thickness and there is a phase
$-1$ associated with each self-linking of a vortex ribbon. We
briefly also discuss a different SPT phase that occurs for $U(1)
\times Z_2^T$ where each vortex loop can be viewed as a $1d$
Haldane spin chain. These results are obtained by analysing both
the sigma model effective field theory and the topological ``BF"
effective field theories proposed in Ref.
\onlinecite{vishwanathsenthil} for these phases.

In $2d$ a result with a similar flavor has been derived by Levin
and Gu\cite{levingu} for an SPT phase with Ising, {\em i.e} $Z_2$,
symmetry in terms of a domain wall loop gas with phase factors. In
Sec. \ref{2d} we reproduce this result using our methods. We also
discuss the ground state wavefunction structure of the 2d boson
topological insulator with  $U(1) \times Z_2^T$ symmetry.  We use
this to obtain a dual vortex description of this state, and show
that the physics is correctly captured.

\begin{figure}
\includegraphics[width=2.9 in]{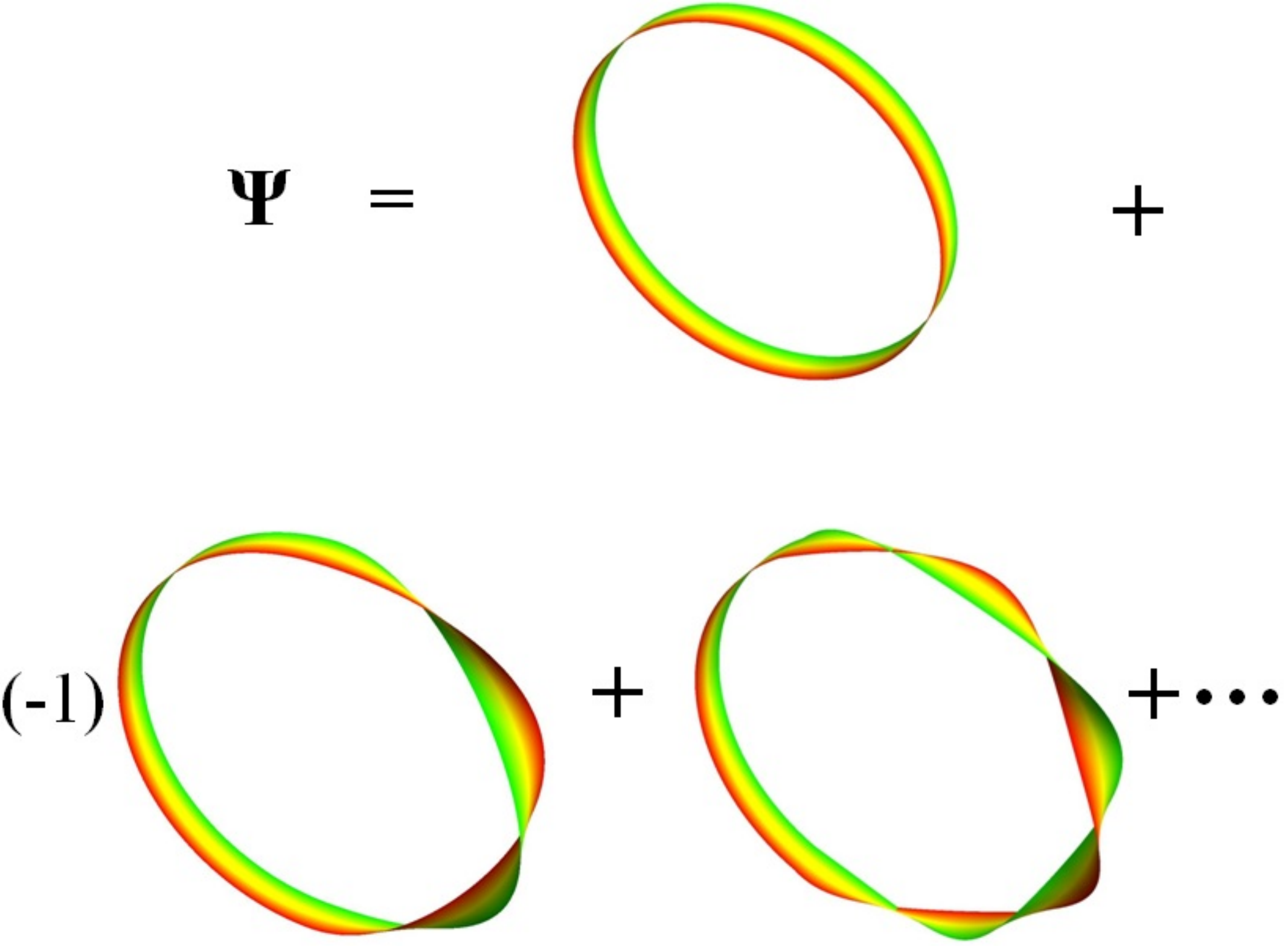}
\caption{ The wave function of the 3d bosonic SPT discussed in
this paper is a superposition of all the configurations of vortex
ribbons with factor $(-1)$ associated with each self-linking.}
\label{ribbon}
\end{figure}

\section{Wave function of trivial 3d Bose Mott insulator}
\label{wftrbmi}

Let us start with briefly reviewing the trivial Mott insulating
phase of bosons. This is conveniently modeled by a quantum
disordered phase of interacting U(1) rotors on a 3d lattice, which
is described by the Hamiltonian $H = \sum_{<i,j>} - t
\cos(\theta_i - \theta_j) + U (\hat{n}_i)^2$. The boson creation
operator $b_i = e^{i\theta_i}$ and $n_i$ is the corresponding
$U(1)$ charge at site $i$. $\theta_i$ and $n_i$ are canonically
conjugate. The quantum disordered phase of the rotors is
equivalent to the familiar Mott insulator phase and occurs when
$t/U \ll 1$. In the strong coupling limit $ t \rightarrow 0$, the
ground state wave function is a trivial direct product state:
\beqn | \Psi \rangle = \prod_i |\hat{n}_i = 0 \rangle \sim \prod_i
\int_0^{2\pi} d\theta_i |\theta_i \rangle. \label{wf0} \eeqn The
wave function of the quantum disordered phase with finite but
small $t/U$ can be derived through perturbation on wave function
Eq.~\ref{wf0}.  For our purposes it is useful to consider a simple
approximate form of the wave function that captures the physics of
the Mott phase : \beqn | \Psi \rangle \sim \int_0^{2\pi} \prod
d\theta \exp[\sum_{<i,j>} - K \cos(\theta_i - \theta_j)] \prod_l
|\theta_l \rangle, \eeqn where $K \sim t/U \ll 1$. This wave
function is a superposition of configurations of $\theta_i$ with a
weight that is the same as the Boltzman weight of the 3d classical
rotor model. The standard duality formalism of the 3d classical
rotor model leads to the dual representation of this wave
function: \beqn | \Psi \rangle &\sim& \int D \vec{A}
\sum_{\vec{J}} \exp[  - \int d^3x \frac{1}{2K}(\vec{\nabla} \times
\vec{A}) ^2 + i 2\pi \vec{A} \cdot \vec{J}] \cr\cr &\times&
|\vec{A}(x), \vec{J}(x) \rangle, \label{wf1} \eeqn 
Vector field $\vec{J}$ takes only
integer values on the dual lattice, and it represents the vortex
loop in the phase $\theta$. In order to guarantee the gauge
invariance of $\vec{A}$, $\vec{J}$ must have no source in the
bulk: $\vec{\nabla} \cdot \vec{J} = 0$. The vortex loop $\vec{J}$
can only end at the boundary, which corresponds to a 2d vortex.
The U(1) gauge field $\vec{A}$ induces long range interactions
between vortex loops with coupling strength $K$. In the limit $K
\rightarrow 0$, $i.e.$ the strong coupling limit of the original
rotor, the wave function Eq.~\ref{wf1} for quantum disordered
lattice bosons becomes a equal weight superposition of all vortex
loop configurations, with a weak long range interaction.

Quite generally the Mott insulating phase is obtained when the
vortex loops have proliferated. Consequently the ground state wave
function can be described as a loop gas of oriented interacting
vortex loops. The discussion above provides a derivation of this
loop gas wave function starting from a simple but approximate
microsopic boson wave function. A crucial point about the
structure of the loop gas wave function for the trivial Mott
insulator is that it has {\it positive} weight for all loop
configurations.

\section{Wave function of 3d Bosonic SPT phases}
\label{wf3dtbi}

A 3d SPT phase with U(1) symmetry is also a quantum disordered
phase of rotor $\theta_i$, thus it is expected that its wave
function is still a superposition of vortex loop configurations.
However, more physics needs to be added to the vortex loops in
order to capture the novel physics of the SPT phase. One of the
central results of this paper is to determine the structure of
this vortex loop gas wave function for the 3d SPT phases with U(1)
and time-reversal symmetry discussed in
Ref.~\onlinecite{vishwanathsenthil}. We first focus on one example
which occurs for both $U(1) \times Z_2^T$ and for $U(1) \rtimes
Z_2^T$. We show that the ground state  is described by a
superposition of vortex loop configurations $|C_v\rangle$, but
each vortex loop should be viewed as a ``ribbon" rather than a
line, and a self-linking of this ribbon contributes exactly factor
$(-1)$ (Fig.~\ref{ribbon}): \beqn | \Psi \rangle \sim \sum_{C_v}
(-1)^{N_t} \psi_0\left[ C_v\right]| C_v \rangle, \label{wfspt3d}
\eeqn where $N_t$ is the number of self-linkings. Here
$\psi_0\left[C_v\right]$ is the weight of that vortex
configuration in a trivial Mott insulator. The self-linking of a
vortex ribbon is the linking number between the loops defined by
the 2 ends of the ribbon.

This wave function Eq.~\ref{wfspt3d} explains the key phenomena of
the 3d SPT phase discussed in Ref.~\onlinecite{vishwanathsenthil}.
In Ref.~\onlinecite{vishwanathsenthil}, using a 2+1d boundary
field theory, it was proved that the vortex of the U(1) rotor at
the boundary of this SPT is a fermion. \footnote{This means that
the dual effective field theory of the surface is in terms of a
fermionic vortex field rather than the usual dual vortex theory
which is in terms of a bosonic vortex}. A vortex at the boundary
is the end (source) of the vortex ribbon in the bulk. As was
discussed in Ref.~\onlinecite{teokane}, and as shown below,
exchanging the ends of ribbons is equivalent to twisting one of
the ribbons by $2\pi$, which according to Eq.~\ref{wfspt3d} should
contribute factor $(-1)$. Thus the bulk wave function
Eq.~\ref{wfspt3d} already implies that the vortex at the boundary
must be a fermion.


The wave function Eq.~\ref{wfspt3d} can be derived either using a
bulk non-linear sigma model effective field theory for the SPT
phase or using a bulk topological ``BF" field theory. We will
present both these derivations below. In Appendix \ref{latticethry} we present a lattice regularized space time path integral for these SPT phases which is equivalent to these
effective field theories, and which may be preferred by some readers.

  In order to describe the 3d
SPT with either $U(1)\rtimes Z_2^T$ or $U(1)\times Z_2^T$
symmetry,  Ref.~\onlinecite{vishwanathsenthil} proposed the
following nonlinear Sigma model that involves a five component
unit vector $\vec{n} = (n^1, \cdots, n^5)$, with a topological
$\Theta-$term at $\Theta = 2\pi$: \beqn S = \int d^3x d\tau \
\frac{1}{g} (\partial_\mu \vec{n})^2 + \frac{i \Theta}{ \Omega_4 }
\epsilon_{abcde} n^a
\partial_x n^b \partial_y n^c
\partial_z n^d
\partial_\tau n^e, \label{o5nlsm} \eeqn where $\Omega_4$ is the volume of a four
dimensional sphere with unit radius. Eq.~\ref{o5nlsm} has an
enlarged SO(5) symmetry, but later we will reduce this symmetry
down to physical $U(1)\rtimes Z_2^T$ or $U(1)\times Z_2^T$.

In 3+1d, an order-disorder phase transition occurs while tuning
$g$. We will focus on the quantum disordered phase with strong
coupling $g$. Since $\Theta = 2\pi$ in Eq.~\ref{o5nlsm}, its
quantum disordered phase has the same bulk spectrum as the case
with $\Theta = 0$. Thus coupling constant $g$ flows to infinity in
the quantum disordered phase and this is the limit we will focus on
in this paper.

The physical meaning of the $\Theta-$term in a NLSM is usually
interpreted as a factor $\exp(i\Theta)$ attached to every
instanton event in the space-time. Then this interpretation would
lead to the conclusion that $\Theta = 2\pi$ is equivalent to
$\Theta = 0$. However, this interpretation is very much
incomplete, because it only tells us that theories with $\Theta =
2\pi$ and $0$ have the same partition function when the system is
defined on a compact manifold. These two theories actually have
very different ground state wave functions. In order to expose the
wave function, we need to keep an open boundary of time. In this
case the wave function can be derived using the following path
integral: \beqn
 && \langle
 \vec{n}(x) | \Psi \rangle \langle \Psi | \vec{n}^\prime (x)
\rangle \cr\cr &\sim& \int D\vec{n}(x, \tau) \exp( -
S)_{\vec{n}_{\tau = + \infty} = \vec{n}^\prime, \vec{n}_{\tau = -
\infty} = \vec{n}}. \eeqn The ground state wave function $| \Psi
\rangle$ can then be obtained straightforwardly in the strong
coupling limit $g \rightarrow +\infty$: \beqn  | \Psi \rangle &
\sim &  \int D\vec{n}(x)   W[{\vec n}] |\vec{n}(x)\rangle \ \ \ \
\ \ \ \ \ \ \  \cr\cr W[{\vec n}] &=  & e^{\frac{i 2\pi}{\Omega_4}
\int d^3x \int_0^1 du \ \epsilon_{abcde} n^a
\partial_x n^b \partial_y n^c \partial_z n^d
\partial_u n^e}.
 \label{wfo53d}\eeqn

Here $\vec{n}(x, u)$ is an extension of the real space
configuration $\vec{n}(x)$ that satisfies $\vec{n}(x, 0) = (0 ,0
,0 ,0, 1)$, and $\vec{n}(x, 1) = \vec{n}(x)$. Eq.~\ref{wfo53d} is
a superposition of all the configurations of the O(5) vector field
$\vec{n}(x)$, with a weight that is proportional to the real space
Wess-Zumino-Witten (WZW) term $W[\vec n]$ at level-1. Thus the
ground state wave function of Eq.~\ref{o5nlsm} with $\Theta =
2\pi$ is fundamentally different from the case with $\Theta = 0$.
A similar relation between the bulk $\Theta-$term of 1+1d O(3)
NLSM and its ground state wave function was discussed previously
in Ref.~\onlinecite{zhang1989}, in the context of 1d spin chain.

Now let us reduce the artificial SO(5) symmetry of
Eq.~\ref{o5nlsm} to $U(1)$ and $Z_2^T$. We decompose the five
component vector $\vec{n}$ as $ \vec{n} = (\sin(\alpha)
\vec{\phi}, \ \cos(\alpha)\phi_0 )$, where $\vec{\phi}$ is a unit
four component vector, and $\phi_0 = \pm 1$ is an Ising order
parameter. We further define two bosonic rotor operators $b_1 \sim
\phi_1 + i \phi_2$, $b_2 \sim \phi_3 + i \phi_4$. Under the U(1)
and $Z_2^T$, we take these variables to transform as
 \beqn
 Z_2^T &:& b_1, b_2 \rightarrow b_1, b_2 \ (U(1) \rtimes Z_2^T), \cr\cr
 && b_1, b_2 \rightarrow - b^\ast_1, -  b^\ast_2 \ (U(1) \times Z_2^T), \cr\cr
 && \phi_0 \rightarrow - \phi_0, \cr\cr
 U(1) &:& b_1 \rightarrow e^{i \theta} b_1, \ \ b_2 \rightarrow e^{i \theta} b_2.
\eeqn We assume the system favors $\vec{\phi}$ over $\phi_0$. If
the time-reversal symmetry is preserved, namely $\langle
\phi_0\rangle = 0$, the WZW term in the wave function
Eq.~\ref{wfo53d} reduces to a theta term for the 4-component unit
vector $\vec \phi$ in $3+0$ dimensions. Thus we get the following
wave function: \beqn |\Psi \rangle &\sim& \int D\vec{\phi}(x)
\cr\cr &\times& \exp( \int d^3x \frac{i\pi}{12\pi^2}
\epsilon_{abcd} \epsilon_{\mu\nu\rho} \phi^a
\partial_\mu \phi^b
\partial_\nu \phi^c \partial_\rho \phi^d) | \vec{\phi}(x) \rangle
\cr\cr & = & \int D\vec{\phi}(x) \ (-1)^{N_s} | \vec{\phi}(x)
\rangle. \label{wfo43d}\eeqn This wave function is a superposition
of all configurations of $\vec{\phi}(x)$ in real space, with a
$\theta-$term defined in 3d real space, at precisely $\theta =
\pi$. $N_s$ is the Skyrmion number of the four component vector
$\vec{\phi}$, since $\pi_3[S^3] = \mathbb{Z}$. The value $\theta =
\pi$ in the wave function is protected by time-reversal symmetry
$Z_2^T$.  If this $Z_2^T$ symmetry is broken, $\theta$ in this
wave function will be tuned away from $\pi$.

\begin{figure}
\includegraphics[width=3.3 in]{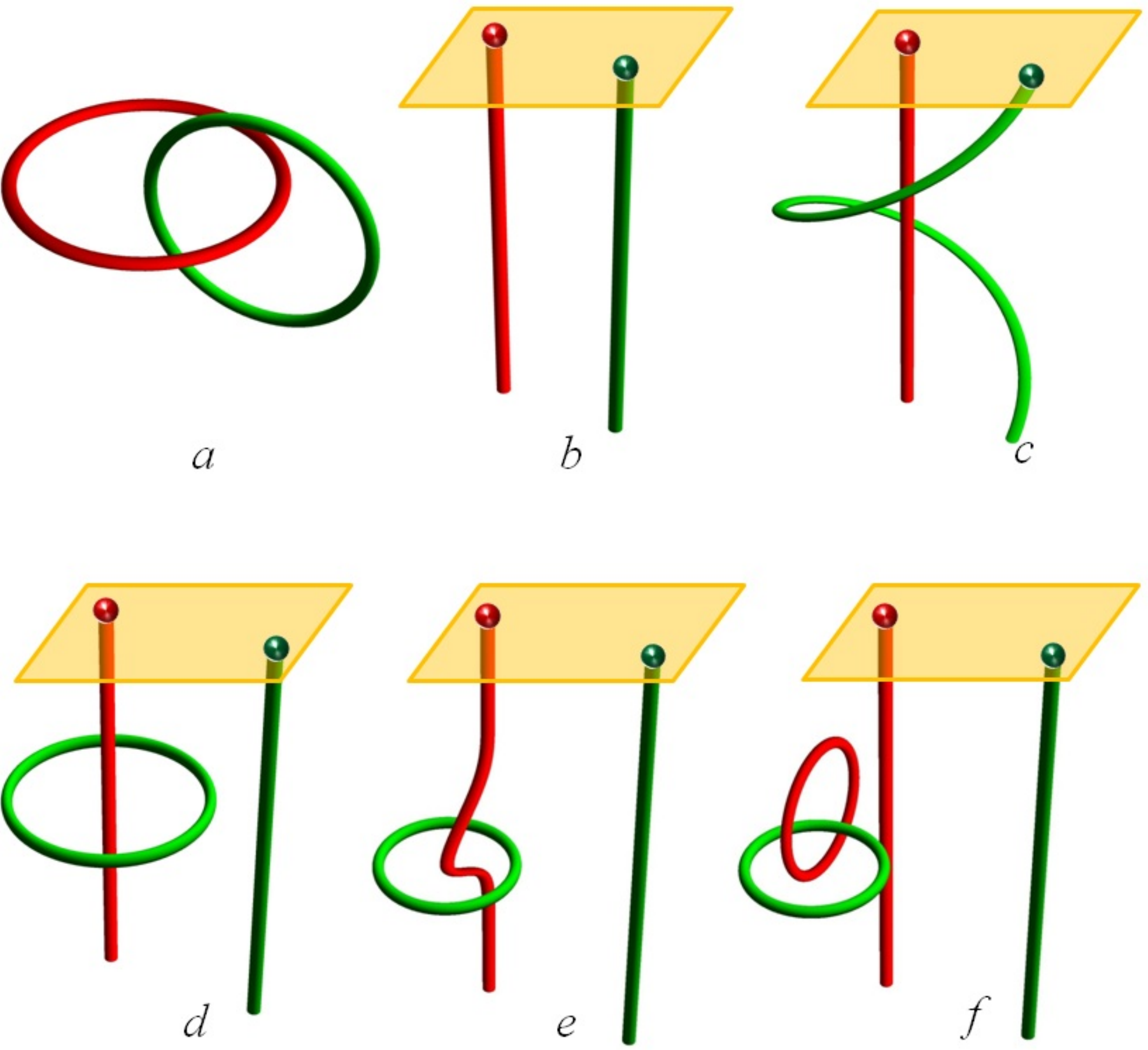}
\caption{$(a)$.
When the symmetry is $U(1) \times U(1)$ the bulk
wave function is a superposition of two flavors of vortex loops
with factor $(-1)$ attached to each linking. $(b-f)$, braiding
between two flavors of vortices at the boundary effectively
creates one extra linking to the bulk vortex loops, which
according to the bulk wave function would contributes factor
$(-1)$. This implies that the two flavors of vortices at the
boundary have mutual semion statistics.}
\label{vortexloop}
\end{figure}

For our purposes we need to introduce anisotropies that reduce the
symmetry from $O(4)$ to $U(1) \times U(1)$.  Then the $\theta$
term (at $\theta = \pi$) implies that there is a phase factor $-1$
each time the vortex loops of the two boson species
link~\cite{senthilfisher}. Thus this two species boson Mott
insulator has a wave function which is a superposition of all
vortex loops of the two species with a crucial factor of $(-1)^L$
where $L$ is the total number of linked opposite species vortex
loops.  In contrast for the trivial Mott insulator of the two
boson species system, the weight for all vortex loop
configurations can be taken to be positive.

It is implicit in the discussion in terms of a four component unit
vector $\vec{\phi}$ that classical configurations of the $b_{1,2}$
fields are always such that $b_{1,2}$ cannot simultaneously
vanish. As the amplitude of either of these fields vanishes in
their vortex core this implies that the vortex loops of the two
species cannot intersect. Thus a configuration with a linking of
the two vortex loops cannot be deformed to one without a linking.

This bulk wave function Eq.~\ref{wfo43d} also implies that at the
2d boundary, the vortex of $b_1$ and vortex of $b_2$ (sources of
vortex loops) have a mutual semion statistics, because when one
flavor of vortex encircles another flavor through a full circle,
the bulk vortex loops effectively acquire one extra linking
(Fig.~\ref{vortexloop}), which according to the bulk wave function
would contribute factor $(-1)$.

Let us now provide an alternate derivation of this result using
the bulk topological BF theory for the SPT phase also proposed in
Ref. \onlinecite{vishwanathsenthil}. This theory takes the form
\begin{equation}
2\pi {\mathcal L}_{3D} =
\sum_I\epsilon^{\mu\nu\lambda\sigma}B^I_{\mu\nu}\partial_\lambda
a^I_\sigma + \Theta \sum_{I,J}\frac{K_{IJ}}{4\pi}
\epsilon^{\mu\nu\lambda\sigma}\partial_\mu a^I_\nu
\partial_\lambda a^J_\sigma \label{BFOverview}
\end{equation}
Here $B^I_{\mu \nu}$ is a rank-$2$ antisymmetric tensor that is
related to the current of boson of species $I = 1,2$ through
$j^I_\mu = \frac{1}{2\pi} \epsilon_{\mu\nu\lambda\sigma}
\partial_\nu B^I_{\lambda\sigma}$.  $a^I_\mu$ is a 1-form gauge
field which describes the vortices of the bosons. Specifically the
magnetic field lines of $a^I$ are identified with the vortex lines
of the boson of species $I$.  For the SPT state of interest the
$K$ matrix is simply $\sigma_x$. The parameter $\Theta = \pi$ (not
to be confused with the theta parameter in the sigma model
description). The crucial difference with the trivial Mott
insulator is the second $\Theta$ term. To get the ground state
wavefunction we again evaluate the Euclidean path integral with
open temporal boundary conditions. Using the well known fact that
the $\Theta$ term is the derivative of a Chern-Simons term we end
up with the following ground state wave functional:
\begin{equation}
\psi\left[ a^I_i, B^{IJ}_{jk}\right] \sim e^{i\frac{\Theta}{8\pi^2}\int
d^3x \epsilon_{ijk} K^{IJ} a^I_i \partial_j a^J_k} \psi_0\left[
a^I_i, B^{IJ}_{jk}\right]
\end{equation}
Here $\psi_0$ is the wave functional for the trivial Mott
insulator. The wave functional for the SPT insulator is thus
modified by a phase factor given by a $3+0$ dimensional
Chern-Simons term. As is well known the Chern-Simons term is
related to a counting of the total linking number of the
configuration of the magnetic flux lines of the gauge fields.
Specializing to the case at hand we see that in the presence of a
$2\pi$ flux line of $a_1$, there is a phase factor $\Theta = \pi$
whenever a $2\pi$ flux line of $a_2$ links with it. Thus we
reproduce the result that there is a phase of $\pi$ associated
with each linking of opposite species vortex lines.


Finally if the $U(1)\times U(1)$ symmetry is broken down to
diagonal U(1), then the vortex loops of the two species will be
confined to each other. The resulting common vortex loop of the
rotor $b \sim b_1 \sim b_2$ becomes a ribbon, whose two edges are
the vortex loops of $b_1$ and $b_2$. Further for simplicity we
assume that there is an energetic constraint at short distances
that prevent two vortex lines of the same species from approaching
each other. In particular we assume that the binding length scale
of the opposite species vortex loops is smaller than the allowed
separation between same species vortex loops. Then the vortex
ribbons cannot intersect each other. Such a ``hard-core"
constraint on the short distance physics should not affect the
universal long distance behavior of the
wavefunction\footnote{Indeed if we do allow intersection of the
ribbons then the phase factors are obtained by going back to the
original $U(1) \times U(1)$ theory with the two species of vortex
loops. The embedding of the $U(1)$ in the higher $U(1) \times
U(1)$ symmetry gives a short distance `regularization' of ribbon
intersections. Banning intersections enables us to discuss the
essential physics without complications.}. Note that the binding
of the two species of vortex loops gives a physical implementation
of the mathematical concept of `framing' used to describe the
topology of knots. The linking between the two flavors of vortex
loops becomes a self-linking of the ribbon. Thus wave function
Eq.~\ref{wfo43d} reduces to wave function Eq.~\ref{wfspt3d}. As we
mentioned before, this bulk wave function Eq.~\ref{wfspt3d}
implies that the end point of a vortex ribbon at a 2d boundary is
a fermion.

Similarly bulk external sources for vortex ribbons will also be
fermions. Such sources are points in three dimensional space where
we force the vortex lines to end. In the original boson Hilbert
space vortex lines cannot of course end. So these bulk vortex
sources are to be regarded as probes of the system where we
locally modify the Hilbert space. The statistics of these bulk
external vortex sources is readily understood from the bulk wave
function and is discussed in Appendix \ref{vrtxsrce}.  A quick
hint of the fermionic statistics comes from asking about the
behavior under $2\pi$ spatial rotations. The vortex ribbon
emanating from a vortex source is twisted by $2\pi$, and this has
an extra phase $-1$ compared to the untwisted ribbon. Thus the
vortex source has `topological' spin$1/2$ as expected if it is a
fermion.

All these results concur with the boundary theory discussed in
Ref.~\onlinecite{vishwanathsenthil}.  There a boundary field
theory for the SPT is derived, which is a 2+1d NLSM with four
component vector $\vec{\phi}$, and there is a 2+1d space-time
$\Theta-$term at precisely $\Theta = \pi$. This space-time
$\Theta-$term implies that the vortex of $b_2$ carries
$1/2-$charge of $b_1$, and vice versa. Thus vortices of $b_1$ and
$b_2$ have a mutual semion statistics. When the symmetry is broken
down to one single U(1), the 2d vortex at the boundary becomes a
bound state of the two flavors of vortices: thus eventually this
bound vortex becomes a fermion.

Ref.~\onlinecite{vishwanathsenthil} also described a different
interesting SPT phase with $U(1) \times Z_2^T$ symmetry. There the
surface theory is such that the surface vortex carries a Kramers
doublet in its core.  A bulk effective field theory of this phase
is also obtained~\cite{vishwanathsenthil} by starting with the
$O(5)$ non-linear sigma model (Eq.~\ref{o5nlsm}) with anisotropies
but with a different realization of symmetry from the one
described above. For instance, we can decompose five component
vector $\vec{n}$ in a different way: $\vec{n} = (\mathrm{Re}[b],
\mathrm{Im}[b], N^x, N^y, N^z)$, where $b$ is a rotor field that
transforms under $Z_2^T$: $b \rightarrow - b^\ast$. $\vec{N}$ is a
three component vector that changes sign under $Z_2^T$ but is
uncharged under the global $U(1)$. The 3+1d $\Theta-$term in
Eq.~\ref{o5nlsm} implies that the vortex loop of $b$ is in a 1d
Haldane phase of vector $\vec{N}$. We may now break the symmetry
down to just $U(1) \times Z_2^T$. Then a vortex at the boundary
must carry a Kramer's doublet, because it is effectively the edge
of the 1d Haldane-like phase (with $Z_2^T$ symmetry) along the
vortex loop. This is the ``Phase 1 SPT" with symmetry $U(1) \times
Z_2^T$ discussed in Ref.~\cite{vishwanathsenthil}.

Let us now consider the ground state wave function for this phase
which in terms of the 5-component vector is still given by Eqn.
\ref{wfo53d}. Now the interpretation of the WZW term is different.
As is familiar from discussions of deconfined quantum criticality
in terms of sigma models with WZW
terms\cite{tanakahu,senthilfisher,groversenthil} in $2+1$
dimensions, with these symmetries an external source for a vortex
line carries spin-$1/2$ of the $O(3)$ rotation that acts on the
$\vec N$ vector. If the $O(3)$ symmetry is broken down to $Z_2^T$,
then the vortex source still has a Kramers doublet. This of course
is completely consistent with the picture that each vortex may be
viewed as  a Haldane chain. Thus the ground state wave function in
this case can be viewed as a vortex loop gas of Haldane chains.

\begin{figure}
\includegraphics[width=2.9 in]{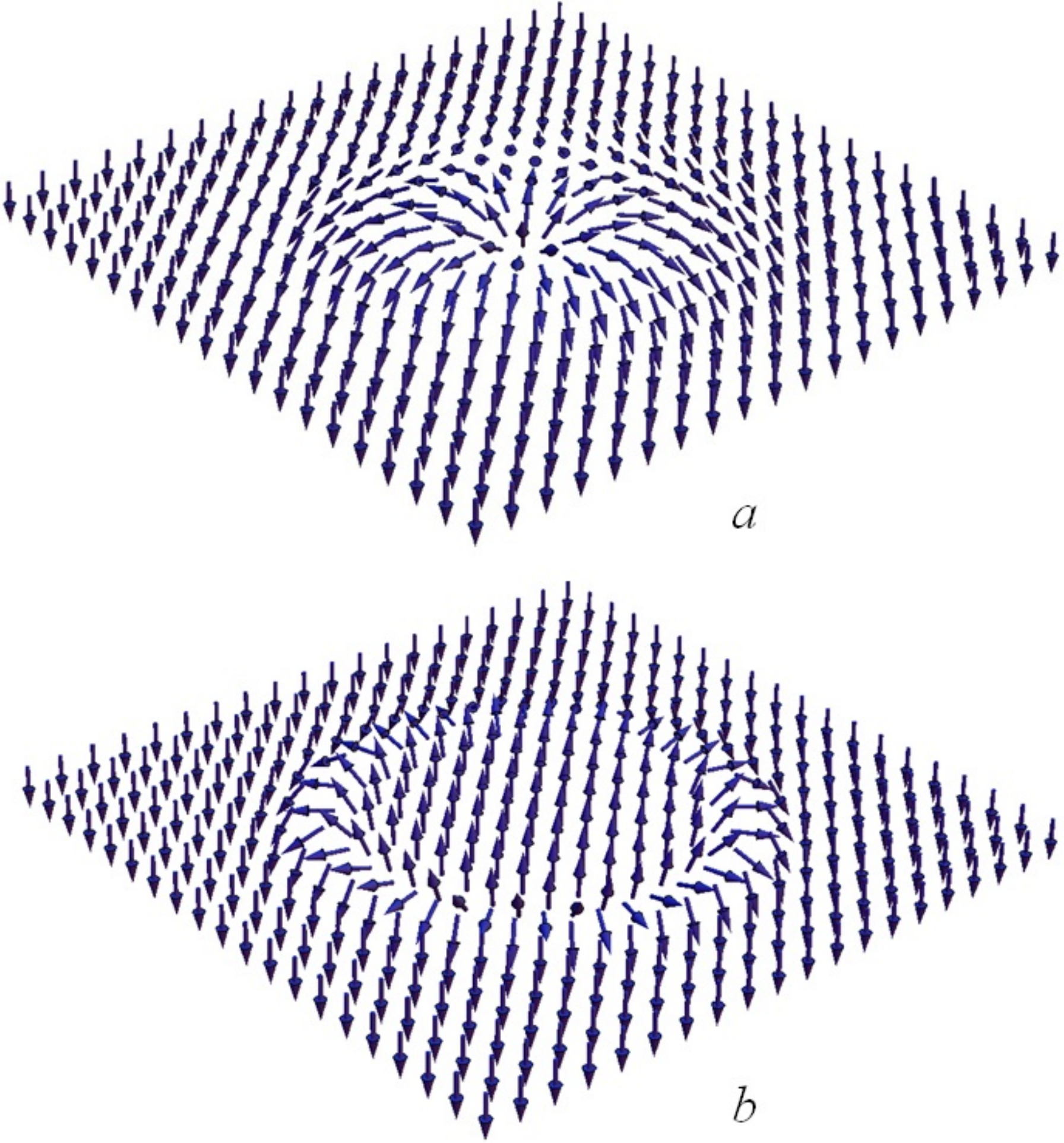}
\caption{$(a)$ Skyrmion of O(3) vector $\vec{n}$ in 2d space.
$(b)$ If the SO(3) symmetry is broken down to $Z_2$, the Skyrmion
becomes a domain wall of $Z_2$ order parameter $n^z$. }
\label{skyrmion}
\end{figure}

\section{Ground state wave function of 2d SPT phases}
\label{2d}

\subsection{2d SPT phase with $Z_2$ symmetry}

Let us now switch gears to SPT phases in 2d. We begin by making
contact the work of Levin and Gu~\cite{levingu} on the Ising SPT
phase. The simplest SPT phase in 2d has a $Z_2$ global symmetry.
In Ref.~\onlinecite{levingu}, a lattice model for this phase has
been discussed. The ground state wave function was argued to be a
sum over all domain wall configurations with a factor $(-1)^{N_d}$
where $N_d$ is the  total number of domain wall loops. We now show
how to reproduce this results within the methods of this paper.


In $2+1$ space-time dimensions many SPT phases are conveniently
described by starting with an effective non-linear sigma model
field theory in terms of a four component unit vector $\vec \phi$.
The field theory action reads \beqn S = \int d^2x d\tau \
\frac{1}{g}(\partial_\mu \vec{\phi})^2 + \frac{i 2\pi}{12\pi^2}
\epsilon_{abcd}\epsilon_{\mu\nu\rho} \phi^a
\partial_\mu \phi^b
\partial_\nu \phi^c \partial_\rho \phi^d,
\label{o4}\eeqn
The crucial ingredient is the $\Theta$ term for the 4-component unit vector
 at $\Theta = 2\pi$. This action
Eq.~\ref{o4} has an SO(4) symmetry, and this SO(4) symmetry
contains a subgroup $Z_2$ symmetry $\vec{\phi} \rightarrow -
\vec{\phi}$.
Eventually we will break the artificial SO(4) symmetry of
Eq.~\ref{o4} down to this $Z_2$ subgroup.

In the paramagnetic phase, $i.e.$ in the limit $g \rightarrow +
\infty$, the bulk ground state wave function is
\beqn && |\Psi
\rangle \sim \int D \vec{\phi}(x) \  \exp\{ \frac{i 2\pi}{12\pi^2}
\ \times \cr\cr && \int d^2x \int_0^1 du \
\epsilon_{abcd}\epsilon_{\mu\nu\rho} \phi^a
\partial_\mu \phi^b
\partial_\nu \phi^c \partial_\rho \phi^d \} \ |\vec{\phi}(x)\rangle,
\label{o4wf} \eeqn
which is a superposition of all configurations
of $\vec{\phi}(x)$ with a real space WZW weight.

Now let us decompose the four component vector into $\vec{\phi} =
(\cos(\alpha)\phi_0, \ \sin(\alpha)\vec{n})$, where $\phi_0 = \pm
1$ is an Ising order parameter, and $\vec{n}$ is a unit three
component vector: $(\vec{n})^2 = 1$.
We also break the SO(4)
symmetry down to $Z_2 \times \mathrm{SO}(3)$ symmetry: \beqn Z_2
&:& \phi_0 \rightarrow - \phi_0, \ \ \ \vec{n} \rightarrow -
\vec{n}, \cr\cr \mathrm{SO}(3) &:& \mathrm{Rotation} \ \mathrm{of}
\ \vec{n}. \eeqn
Under this symmetry reduction, if the system
energetically favors vector $\vec{n}$ over $\phi_0$ (favors
$\alpha = \pi/2$), the wave function Eq.~\ref{o4wf} reduces to
\beqn |\Psi \rangle &\sim& \int D \vec{n}(x) \ \exp\{ \frac{i\pi
}{8\pi} \int d^2x \ \epsilon_{abc}\epsilon_{\mu\nu} n^a
\partial_\mu n^b \partial_\nu n^c \} \ |\vec{n}(x)\rangle
\cr\cr &=& \sum_{N_s} (-1)^{N_s} | \vec{n}(x) \rangle,
\label{o3wf} \eeqn where $N_s$ is the number of Skyrmions of O(3)
vector $\vec{n}$ in the 2d space. As long as we keep the $Z_2$
symmetry $\vec{\phi} \rightarrow - \vec{\phi}$, the expectation
value of $\phi_0$ is zero, then each Skyrmion will contribute a
phase factor exactly $(-1)$.


Now let us further break $Z_2 \times \mathrm{SO}(3)$ symmetry down
to $Z_2 \times \mathrm{SO}(2)$: \beqn Z_2 &:& \phi_0 \rightarrow -
\phi_0, \ \ \ \vec{n} \rightarrow - \vec{n}, \cr \cr
\mathrm{SO}(2) &:& \mathrm{Rotation} \ \mathrm{of} \ n^x, \ n^y.
\eeqn Also we assume that the system favors $n^z$ over $n^x$ and
$n^y$, then each skyrmion becomes a domain wall of $Z_2$ order
parameter $n^z(x)$ (Fig.~\ref{skyrmion}). Now the wave function
Eq.~\ref{o3wf} reduces to a superposition of configurations of
$Z_2$ order parameter $n^z(x)$: \beqn | \Psi \rangle \sim
\sum_{n_z} (-1)^{N_d}| n^z(x)\rangle, \label{z2wf} \eeqn where
$N_d$ is the number of closed domain wall loops of  $n^z(x)$.
Eventually we can also break the residual SO(2) symmetry, and the
wave function Eq.~\ref{z2wf} is unchanged.

The wave function Eq.~\ref{z2wf} is exactly the one derived from
the lattice model of 2d SPT phase with $Z_2$
symmetry~\cite{levingu}. In the appendix we will also demonstrate
that the effective field theory Eq.~\ref{o4} implies that, after
coupling this SPT phase to a dynamical $Z_2$ gauge field, the
$\pi-$flux of this $Z_2$ gauge field has a semion statistics,
which is consistent with the result in Ref.~\onlinecite{levingu}.


With a full SO(4) symmetry, the edge states of Eq.~\ref{o4} with
precisely $\Theta = 2\pi$ is the nonchiral SU(2)$_1$ conformal
field theory (or equivalently as an SO(4) non-linear sigma model
with a level-$1$ WZW term). Since the original SO(4) symmetry is
reduced to its $Z_2$ subgroup $\vec{\phi} \rightarrow -
\vec{\phi}$, we have to argue that the edge state of Eq.~\ref{o4}
survives under this symmetry reduction.   Because the $Z_2$
symmetry acts on all four components of $\vec{\phi}$, in the
boundary WZW model, terms allowed by the $Z_2$ symmetry are
$\sum_{i,j }g_{ij} \phi_i \phi_j$ ($i, j = 0 ,1,2, 3$).  If these
terms are relevant, it leads to spontaneous $Z_2$ symmetry
breaking and two fold degeneracy at the boundary. Thus the edge
state cannot be completely trivial (gapped and nondegenerate) as
long as the $Z_2$ symmetry is preserved.


\subsection{2d SPT phase with $U(1) \rtimes Z_2^T$ symmetry}

Finally we consider the 2d bosonic topological insulator which
occurs when the global symmetry is $U(1) \rtimes Z_2^T$.  This may
be described by starting again with the same 4-component
non-linear sigma model but with the following implementation of
the physical symmetry. We write $\phi_2 - i\phi_3 = b$ and let $b$
have charge-1 under the global $U(1$). Under $Z_2^T$, we demand $b
\rightarrow b, \phi_0 \rightarrow - \phi_0, \phi_1 \rightarrow
-\phi_1$.  As before we again assume first an anisotropy that
prefers $\vec n$ over $\phi_0$ so that the ground state wave
function is given by Eqn. \ref{o3wf}. Now we introduce further
anisotropy to reduce to the desired $U(1) \rtimes Z_2^T$. The
defects of the charged field $b$ are of course point vortices. In
the core of these vortices the amplitude of $b$ is suppressed and
the $\vec{\phi}$ points entirely in the $\phi_1$ direction. There
are two different vortices - known as merons - depending on
whether in the core $\phi_1  = \pm 1$.  Each meron may be viewed
as half a skyrmion and has $N_s = \mathrm{sgn}{(\phi_1)}1/2$. Thus
the ground state wave function is then a sum over all possible
configurations of the two kinds of meron vortices with phase
factors $e^{\pm i\frac{\pi}{2}}$ for the two kinds of vortex. Let
$n_{v\pm}$ be the vortex number of either species at site $i$ in a
lattice description. Then we require that the total vorticity $N_v
= \sum_i n_{v+} + n_{v-} = 0$. Then the phase factor in the wave
function $e^{i\frac{\pi}{2}\sum_i (n_{v+} - n_{v-})}= (-1)^{\sum_i
n_{v-}}$. Thus there is a relative phase of $-1$ associated with
$-$ vortices compared to $+$ vortices.

We now argue that this structure of the wave function matches what
is known about the 2d bosonic topological insulator. Consider a
dual description of such an insulator. From our arguments above
there are two kinds of vortex fields $\Phi_{v\pm}$ corresponding
to the two meron vortices. The dual vortex theory will have a
Lagrangian
\begin{equation}
{\cal L}_d = \sum_{s = \pm} |(\partial_\mu - ia_\mu)\Phi_{vs}|^2 +
...+ \frac{\kappa}{2} \left(\epsilon_{\mu\nu\lambda} \partial_\nu
a_\lambda \right)^2
\end{equation}
Here $a_\mu$ is the usual fluctuating non-compact $U(1)$ gauge
field whose flux density is the original boson number.  There must
in addition be terms where the meron cores tunnel into each other:
\begin{equation}
\lambda\left(\Phi_{v+}^\dagger \Phi_{v-} + h.c \right)
\label{merontunn}
\end{equation}
We begin by ignoring these and we will reinstate them later.
Under time reversal a vortex must go to
an anti vortex (as the boson phase is odd) and the meron cores
flip into each other. Thus under $Z_2^T$
\begin{equation}
\Phi_{v+} \rightarrow \pm \Phi_{v-}^\dagger
\end{equation}
In the trivial insulator all vortex configurations contribute with
the same sign and we must choose $\Phi_{v+} \rightarrow +
\Phi_{v-}^\dagger$. But for the topological insulator there is a
relative $-$ sign between the two vortex species. Thus we must
choose $\Phi_{v+} \rightarrow - \Phi_{v-}^\dagger$. Condensing
vortices that transform in this manner will give us the boson
topological insulator. Now let us include the meron core tunneling
term. Then the two vortex species mix with each other so that we
identify a single vortex $\Phi_v = \Phi_{v+} \sim \Phi_{v-}^*$.
Its transformation under time reversal is $\Phi_v \rightarrow \pm
\Phi_v^\dagger$ where the $+$ sign describes the trivial insulator
and the $-$ the topological insulator. This is exactly the same
transformation law for the vortices that is dictated by the edge
theory analysis of the 2d boson topological
insulator\cite{levinstern,luashvin}. Thus the wave function
description we developed captures the physics of this state, and
further gives a bulk dual vortex description.


\section{Discussion}

In summary, we have demonstrated in this work that, although most
of the novel phenomena of a SPT phase occur at its boundary, its
bulk ground state wave function is indeed drastically different
from a trivial direct product disordered phase. This bulk wave
function can be conveniently derived from the effective field
theory of the SPT phase. The structure of the ground state wave
functions in terms of dual vortex configurations derived in this
work  provide a simple physical picture of the phenomena
associated with these SPT phases.

The dual ground state bulk wave function provides a nice intuitive
understanding of the differences between ordinary and topological
boson insulators. However in this paper we haven't attempted to
explicitly construct microscopic models for these phases. Progress
in this direction is reported very recently in Refs.
\onlinecite{chongts,burnell} which appeared this paper was
submitted.

Finally  we note that for the 3d boson topological insulators,
there is some superficial similarity with the wave functions of
Walker and Wang~\cite{wwmodel} though a detailed understanding of
the relationship is presently not clear to us. The Walker-Wang
models also have ground state wave functions  as string net
configurations with amplitude determined by a 2+1-d topological
quantum field theory.  In some cases these can correspond to SPT
phases (see Ref.~\onlinecite{burnell}). However it is not clear
how the strings in the Walker-Wang models   are related to the
physical bosons; in particular they are not to be identified with
physical vortex loops. Exploring the connections between the wave
functions in these Walker-Wang constructions of SPT phases and our
dual wave functions is an interesting avenue for future research.

TS thanks A. Vishwanath for an earlier collaboration and
discussions which enabled the present work. We both thank Matthew
Fisher for stimulating discussions. CX was supported by the Alfred
P. Sloan Foundation, the David and Lucile Packard Foundation,
Hellman Family Foundation, and NSF Grant No. DMR-1151208. TS was
supported by NSF DMR-1005434. TS was also partially supported by
the Simons Foundation by award number 229736. He thanks the
Physics Department at Harvard for hospitality where part of this
work was done. We both thank the Institute of Advanced Study at
the Hong Kong University of Science and Technology where this work
was initiated.

\bibliography{wave}

\appendix{}

\section{Appendix: Dynamical $Z_2$ gauge fields in the 2d Ising SPT}

In this appendix we demonstrate that the effective field theory
Eq.~\ref{o4} not only gives us the correct ground state wave
function (Eq.~\ref{z2wf}) of the 2d Ising SPT phase, after
coupling the SPT phase to a $Z_2$ gauge field, the topological
$\Theta-$term of Eq.~\ref{o4} also leads to nontrivial statistics
of the dynamical $\pi-$flux (vison) of the $Z_2$ gauge field.

First of all, Eq.~\ref{o4} can be rewritten as a SU(2) principle
chiral model, by introducing SU(2) matrix field $G = \phi^0
\sigma^0  + i \vec{\phi} \cdot \vec{\sigma} $. $G$ has
SU(2)$-$left and SU(2)$-$right transformations: $G \rightarrow
V^\dagger_L G V_R$. Let us ``gauge" SU(2)$-$left and SU(2)$-$right
transformations with dynamical U(1) gauge fields $a_\mu \sigma^z$
and $b_\mu \sigma^z$, $i.e.$ replace $\partial_\mu G$ with
$\partial_\mu G + i a_\mu \sigma^z G + i b_\mu G \sigma^z$.
According to Ref.~\onlinecite{liuwen,levinsenthil}, after
integrating out matrix field $G$, gauge fields $a_\mu$ and $b_\mu$
both acquire a Chern-Simons term:  \beqn S_{cs} = \int d^2x d\tau
\ \frac{i2}{4\pi} \epsilon_{\mu\nu\rho} a_\mu
\partial_\nu a_{\rho} - \frac{i2}{4\pi} \epsilon_{\mu\nu\rho} b_\mu
\partial_\nu b_{\rho}. \label{cs}\eeqn This is because Eq.~\ref{o4}
also describes a U(1) bosonic SPT with Hall conductivity 2.


A dynamical U(1) gauge field with level$-k$ has the following
properties: its charged quasiparticle carries gauge flux $2\pi/k$,
and this quasiparticle has a statistics angle $\pi/k$. Thus the
Chern-Simons action Eq.~\ref{cs} gives the $\pi-$flux of U(1)
gauge field $a_\mu$ and $b_\mu$ a semion statistics, with
statistics angle $+\pi/2$ and $-\pi/2$ respectively. Notice that
the two U(1) gauge groups share the same $Z_2$ transformation $G
\rightarrow -G$, thus we can break the two U(1) gauge fields down
to one $Z_2$ gauge field, then the dynamical $\pi-$flux of this
$Z_2$ gauge field has two different flavors with semionic
statistics angle $+\pi/2$ and $-\pi/2$ respectively.

In Ref.~\onlinecite{levingu}, using their lattice model, the
authors concluded that the dynamical $\pi-$flux of this $Z_2$
gauge field has a semion statistics. Here we have derived the same
result using our field theory Eq.~\ref{o4}.

\section{Lattice version of effective field theory}
\label{latticethry}

In this Appendix we briefly discuss a lattice regularized version
of the bulk effective field theory for the $3d$ SPT phases
discussed in this paper. We will also briefly review the
considerations leading to the bulk effective `BF + FF' effective
theory of Ref. \onlinecite{vishwanathsenthil}.  This enables a
further elaboration of the discussion in the main text on the
properties of external sources of bulk vortex lines.

To set the stage first consider  the Euclidean lattice action for
3+1-D XY model in Villain form:
\begin{equation}
{\cal S_0} = \sum_{r\mu} \frac{g}{2} \left(j_\mu \right)^2 +
ij_\mu \left(\nabla_\mu \theta_r - 2\pi m_\mu \right)
\end{equation}
Here $m_\mu$ is an integer defined on the links. Physically it
defines the integer vortex current
\begin{equation}
J_{\mu\nu} = \epsilon_{\mu\nu\lambda \kappa}\nabla_\lambda
m_\kappa
\end{equation}
It is slightly more convenient to go to a `gauge' where we
explicitly sum over $m_0$ which has the effect of forcing $j_0$ to
be an integer. We can then drop $m_0$ from the action. Let us also
define $a_\mu = 2\pi m_\mu$. Now consider two species of bosons,
i.e 2 XY models . We put one XY model on a 4d cubic lattice and
the other on a different 4d cubic lattice such that the spatial
links of one lattice penetrate the spatial plaquettes of the
other. (This different treatment of time and space is not
necessary but it helps to visualize). Formally the lattice sites
of one lattice are $(l_0, l_x, l_y, l_z)$ with $l_\mu$ = integer
and  for the other lattice we have $(l_0, l_x + \frac{1}{2}, l_y +
\frac{1}{2}, l_z + \frac{1}{2})$. Then let us write the action for
the coupled XY models:
\begin{eqnarray}
\label{latticeS}
S & = & \sum_{I = 1,2}S_{0 I}+ S_{top} \\
S_{top} & = & i\frac{\Theta}{8\pi^2} K_{IJ}
\epsilon_{\mu\nu\lambda\kappa} \nabla_\mu a_\nu^I \nabla_\lambda
a_\kappa^J
\end{eqnarray}
with the matrix $K = \sigma_x$. The first term $S_{0I}$ is just
the sum of the above XY actions for the each species. The second
`topological' term enforces the phases  associated with the vortex
world sheet configurations.  In the trivial boson insulator
$\Theta = 0$ while in the boson SPT phase $\Theta = \pi$. Thus the
Boltzmann weight for the SPT phase differs from that of the
trivial insulator only through phase factors that depend on the
vortex world sheet configurations.

The action in Eqn. \ref{latticeS}  can be regarded as a lattice
version of the continuum non-linear sigma model used in the main
paper.  Indeed the term $S_{top}$ correctly captures the physics
of the theta term of the sigma model. We now show the relation to
the `BF + FF' effective theory.  As usual a dual description of
the boson system is obtained by writing the conserved boson
$4$-currents in terms of  dual $2$-form fields $B^I_{\mu\nu}$:
\begin{equation}
j^I_\mu = \frac{1}{2\pi}\epsilon_{\mu\nu\kappa\lambda} \nabla_\nu
B^I_{\kappa \lambda}
\end{equation}
The term $j_\mu a_\mu$ in the lattice action above then leads to
the familiar $BF$ term. The integer constraint on $m_\nu =
\frac{a_\mu}{2\pi}$ can be implemented softly by including a term
$-\lambda \cos \left(a^I_\mu - \nabla_\mu \theta^I \right)$. The
$\theta^I$ is just the original boson phase. As explained in Ref.
\onlinecite{vishwanathsenthil}  in the Mott insulator at energies
below the boson gap the bosons may be integrated out to leave
behind just a Maxwell term for the $a_\mu$ fields. The term
$S_{top}$ in the lattice action Eqn. \ref{latticeS} simply becomes
the $FF$ term of the `BF + FF' action.

The properties of external sources of vortex lines can be readily
discussed in terms of these bulk effective field theories. Since
the difference with the trivial insulator comes entirely from the
$S_{top}$ term, it is appropriate to focus first on the vortex
world sheet configurations, take care of the consequences of
$S_{top}$ and then put back the interaction with the smooth part
of the boson phase fields represented by the coupling to
$B^I_{\mu\nu}$.  External bulk sources of vortex lines are simply
monopole sources of $2\pi$ magnetic flux of the internal gauge
fields $a_\mu^I$.  We may now specialize to the SPT phase for
$U(1) \times Z_2^T$ or $U(1) \rtimes Z_2^T$ discussed in the main
text where the boundary vortex is a fermion. Now consider letting
the two boson species tunnel into each other so that the
corresponding vortex lines are bound together to form the vortex
ribbon. Formally in a coarse grained description this is obtained
by setting $a_{1\mu} \sim a_{2\mu}$. Bulk external sources of the
vortex ribbon are simply monopoles of this common internal gauge
field. The wave function based discussion of this paper shows that
these vortex sources are fermions.  Alternately this follows from
the observation of Ref. \onlinecite{vishwanathsenthil} that the
vortex field of the boundary dual vortex theory is a  fermion.

\begin{figure}

\includegraphics[width=3.2 in]{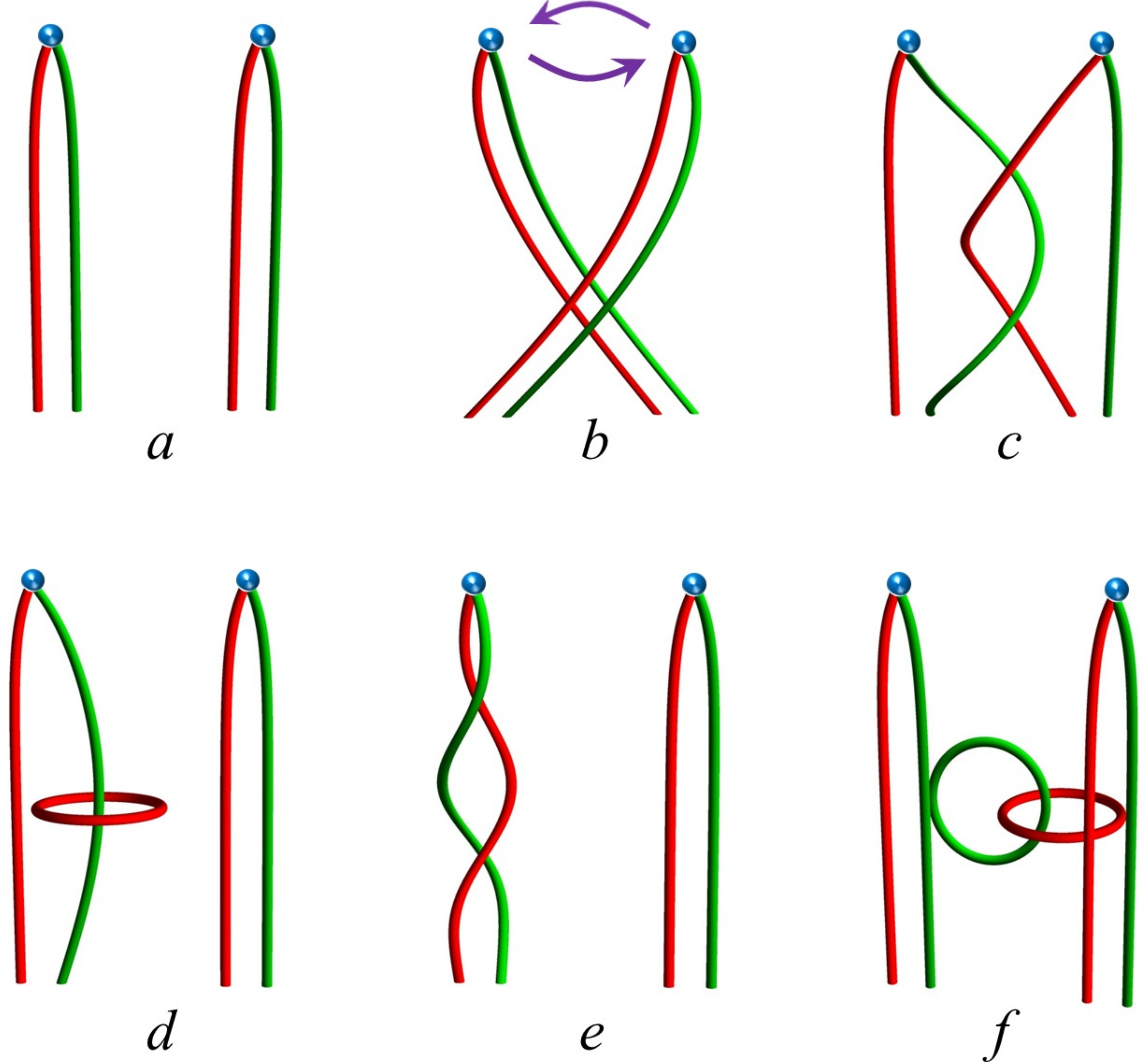}
\caption{$(a)$, a vortex source is the end point of a bulk vortex
ribbon. $(b) \rightarrow (e)$, sequence of vortex ribbon
deformation, starting with interchanging two vortex sources in
$(b)$. $(b)$ is homotopically equivalent to self-twisting one of
the two ribbons by $2\pi$ in $(e)$, which according to the bulk
wave function should acquire factor $-1$. $(f)$, interchanging two
vortex sources is also homotopically equivalent to creating one
extra vortex ribbon with $2\pi$ self-twist in the bulk. }
\label{exchange}
\end{figure}

\section{Statistics of vortex sources}
\label{vrtxsrce}

According to the main text, a vortex at the 2d boundary of the
$U(1) \rtimes Z_2^T$ SPT phase is a fermion. Here we consider bulk
vortex sources. Consider the vortex loop left behind by creating a
vortex source and anti vortex source together, moving the vortex
source around and then annihilating with the anti vortex source.
Now consider creating another such loop. We can envisage two
situations depending on whether or not the two vortex sources were
exchanged with each other during the process of forming the two
closed loops. Pictures corresponding to these two situations may
be found in Fig. 2 of the well known paper by Wilczek and
Zee\cite{zeewilczek}. As argued in that paper the difference
between the two pictures corresponds to a self-linking of one of
the ribbons by $2\pi$. This means that the process of exchange of
the vortex sources has introduced a phase $-1$, and thus the
vortex sources are fermions.

Here we provide further pictures to illustrate this in
Fig.~\ref{exchange}. Let us consider two vortex sources in the
bulk (Fig.~\ref{exchange}$a$). After interchanging two vortex
sources (Fig.~\ref{exchange}$b$), the vortex ribbon configurations
can be continuously deformed into Fig.~\ref{exchange}$e$ ($b
\rightarrow c \rightarrow d \rightarrow e$), which is simply
self-twisting one of the ribbons by $2\pi$. In
Fig.~\ref{exchange}$b$, the ribbon (red and green vortex lines)
connecting to the right vortex source is on the top. Step 1: $(b)
\rightarrow (c)$, connect the two red (green) lines in $(b)$, and
reopen in the horizontal direction; Step 2: $(c) \rightarrow (d)$,
deform the red line on the right into a circle and a straight
line; $(d) \rightarrow (e)$, reconnect the red circle to the red
line on the left, now the ribbon on the left has a $2\pi$
self-twist.

Fig.~\ref{exchange}$b$ can also be continuously deformed into
Fig.~\ref{exchange}$f$, which compared with Fig.~\ref{exchange}$a$
has created another ribbon with a $2\pi$ self-twist, or
equivalently two different vortex loops with linking number 1.
Both configurations $(e)$ and $(f)$ introduce factor $(-1)$
compared with $(a)$.


\bibliography{wave}

\end{document}